\documentclass[twocolumn,showpacs,showkeys,amssymb,nobibnotes,superscriptaddress,aps,pra]{revtex4-1}
\usepackage{amssymb}
\usepackage{amsmath}
\usepackage{amsfonts}
\usepackage{bm}
\usepackage{graphicx}
\usepackage[dvipsnames]{xcolor}
\usepackage{calc}
\usepackage{epsfig}
\usepackage{color}
\usepackage[colorlinks,citecolor=blue]{hyperref}
\begin{document}

\title{Nonlinear effects in modulated quantum optomechanics}

\author{Tai-Shuang Yin}
\author{Xin-You L\"{u}}
\email{xinyoulu@hust.edu.cn}
\affiliation{School of physics, Huazhong University of Science and Technology, Wuhan 430074, China}
\author{Li-Li Zheng}
\affiliation{School of physics, Huazhong University of Science and Technology, Wuhan 430074, China}
\author{Mei Wang}
\affiliation{School of physics, Huazhong University of Science and Technology, Wuhan 430074, China}
\author{Sha Li}
\affiliation{School of physics, Huazhong University of Science and Technology, Wuhan 430074, China}
\author{Ying Wu}
\email{yingwu2@126.com}
\affiliation{School of physics, Huazhong University of Science and Technology, Wuhan 430074, China}
\date{\today}

\begin{abstract}
The nonlinear quantum regime is crucial for implementing interesting quantum effects, which have wide applications in modern quantum science.
Here we propose an effective method to reach the nonlinear quantum regime in a modulated optomechanical system (OMS), which is originally in the weak-coupling regime.
The mechanical spring constant and optomechanical interaction are modulated periodically.
This leads to the result that the resonant optomechanical interaction can be effectively enhanced into the single-photon strong-coupling regime by the modulation-induced mechanical parametric amplification.
Moreover, the amplified phonon noise can be suppressed completely by introducing a squeezed vacuum reservoir, which ultimately leads to the realization of photon blockade in a weakly coupled OMS.
The reached nonlinear quantum regime also allows us to engineer the nonclassical states (e.g., Schr\"{o}dinger cat states) of cavity field, which are robust against the phonon noise.
This work offers an alternative approach to enhance the quantum nonlinearity of an OMS, which should expand the applications of cavity optomechanics in the quantum realm.
\end{abstract}

\maketitle

\section{Introduction}
The last decade has witnessed dramatic progress in the field of cavity optomechanics, exploring the nonlinear interaction between a mechanical oscillator and an optical cavity via radiation pressure force~\cite{Review1,Review2,Sun2015}. For example, considerable achievements have been achieved including cooling the mechanical modes to their quantum ground state~\cite{O’Connell2010,Chan2011Teufel2011,Clark2017},
the observations of normal-mode splitting~\cite{Groblacher2009Nature,Teufel2011}, optomechanically induced transparency~\cite{Weis2010,Safavi-Naeini2011,Karuza2013}, the coherent-state conversion between cavity and mechanical modes~\cite{Fiore2011,Zhou2013,Palomaki2013Palomaki2013}, and a generation of squeezed light~\cite{Brooks2012,Safavi-Naeini2013,Purdy2013}.
Note that the above achievements are realized in the strong driven optomechanical system (OMS), in which the optomechanical interactions are enhanced by a factor $\sqrt{n}$ ($n$ is the mean photon number in the cavity) at the cost of linearizing original radiation-pressure coupling. The reason is the fact that the currently attainable optomechanical radiation-pressure coupling is much smaller than the cavity decay rate~\cite{Groblacher2009Naturephysis,Rocheleau2010,Verhagen2012,Borkje2013,Wu2015}, which limits the OMS to reach a nonlinear quantum regime.

To reach the nonlinear quantum regime, it is highly desirable to realize strong optomechanical radiation-pressure coupling, where the optomechanical coupling strength at the single-photon level exceeds the optical cavity decay rate. Importantly, many interesting nonlinear effects might be demonstrated in this regime, such as macroscopic nonclassical states~\cite{Mancini1997,Bose1997,Marshall2003,Stannigel2012,Komar2013}, the photon blockade phenomenon~\cite{Rabl2011,Kronwald2013Ludwig,Liao2013Law,Lv2013Zhang,Wang2015}, and multiphonon sideband effects~\cite{Nunnenkamp2011}. However, until now the nonlinear quantum regime of an OMS is still a challenging topic, while many theoretical schemes have been proposed to reach the nonlinear quantum regime with current experimental technologies, such as enhancing the radiation-pressure coupling by employing the Josephson effects in superconducting circuits~\cite{Heikkila2014,Johansson2014,Rimberg2014}, the optical coalescence effects~\cite{Genes2013}, and the squeezing effects of the cavity mode~\cite{Lv2015Wu}.

Here, we consider a modulated optomechanical system, which could effectively enter into the nonlinear quantum regime.
Specifically, a mechanical parametric amplification is obtained by the period modulation of the mechanical spring constant, which is experimentally feasible~\cite{Rugar1991,Szorkovszky2011Doherty,Szorkovszky2013Brawley}.
Together with the sinusoidal modulation of the optomechanical coupling, the resonant nonlinear photon-phonon interaction could be enhanced to the single-photon strong-coupling regime by mechanical parametric amplification.
Note that the mechanical parametric amplification could also be used to boost mechanical signals in micro- and nanoelectromechanical systems~\cite{Unterreithmeier2009,Suh2010}, to squeeze the mechanical oscillator beyond the 3dB limit~\cite{Lv2015Liao}, to enhance the photon-photon interaction in a two mode OMS~\cite{Lemonde2016}, to yield optical amplification and squeezing~\cite{Levitan2016}, and to achieve optical non-reciprocity~\cite{Xiong2015,Si2017}.

Moreover, we also show that the phonon-amplification-induced thermal noise can be suppressed by introducing a squeezed vacuum bath of phonon~\cite{Gao2016} with a reference phase matching the phase of parametric amplification. To decrease the amplified phonon noise, other alternative strategies could also be employed, such as adding an additional optical mode to the system~\cite{Rabl2004Shnirman,DallaTorre2013,Tan2013,Kronwald2013}, as well as the so-called "transitionless" driving (TD) protocols~\cite{Lemonde2016}. Under the conditions of enhanced radiation-pressure interaction and suppressed phonon noise, photon blockade and Schr\"{o}dinger cat states could be demonstrated even in an original weakly coupled OMS. In particular, a high fidelity of Schr\"{o}dinger cat states could also be obtained when the amplified phonon noise is not completely suppressed.

Comparing with the previous proposals employing the mechanical parametric pumping~\cite{Lemonde2016,Levitan2016},
the main novelty of our proposal is the realization of dramatically enhancing the resonant nonlinear photon-phonon interaction by associating the mechanical squeezing effects with the modulated optomechanical interaction.
It is quite different from the previous studies without employing the modulated effects of optomechanical interaction~\cite{Lemonde2016,Levitan2016}. Specifically, we consider a single-mode optomechanical system and focus on the regime where the mechanical spring constant and optomechanical interaction are modulated periodically. This modulation generates a resonant nonlinear photon-phonon interaction, which can be enhanced to the single-photon strong-coupling regime. Without employing the modulation effects, the previous study in Ref.~\cite{Lemonde2016} considered a two-cavity OMS and realized the effective enhancement of the photon-photon interaction by the strong mechanical parametric amplification. Our work is also quite different from the work in Ref.~\cite{Levitan2016}, which utilized the coherent two-phonon driving and red mechanical sideband driving applied to the single-mode optomechanical cavity. Thus optical squeezing and amplification could be generated in the linearized regime, but the nonlinear quantum regime was not involved.
\begin{figure}[b]
\centering\includegraphics[width=0.43\textwidth]{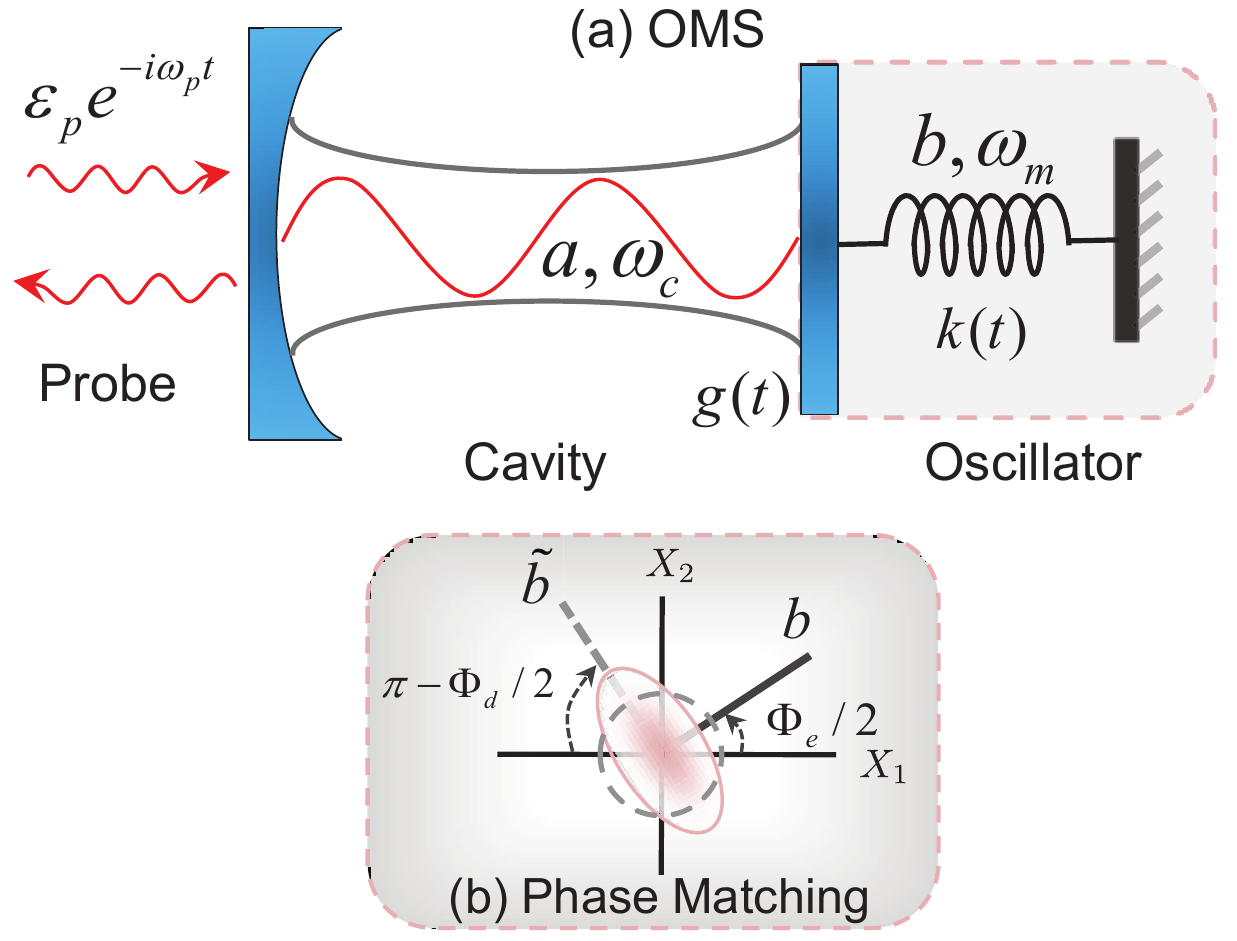}
\caption{(Color online) (a) Schematic illustration of an OMS consisting of an optical cavity mode $a$ coupled to the mechanical mode $b$ with the modulated coupling strength $g(t)=g_{0}\cos(\omega_{d} t).$
In addition, the mechanical spring constant is modulated with frequency $2\omega_d$ and phase $\Phi_{d}$, i.e., $k(t)=k_0 - k_r\cos(2\omega_dt-\Phi_{d})$. A weak probe field with frequency $\omega_p$ and amplitude $\varepsilon_p$ is applied into the cavity to demonstrate the nonlinear quantum regime. (b) The phase-matching condition $\Phi_e-\Phi_d=\pm n\pi$ $( n=1,3,5,... )$ for suppressing the noise of $\tilde{b}$ induced by mechanical parametric amplification.}
\label{fig1}
\end{figure}

This paper is organized as follows: In Sec. II, we introduce the modulated optomechanical system, and the effective Hamiltonian is derived by using the corresponding squeezing transformation and rotating-wave approximation (RWA).
It clearly shows that the nonlinear quantum regime could be reached in our proposal. In Sec. III, we discuss the photon statistical properties featured by the second-order correlation function to demonstrate the nonlinear quantum regime. In Sec. IV, we discuss the generation of nonclassical states of the cavity field to show the application of nonlinear quantum regime.
In Sec. V, we discuss the experimental prospect of our proposal.
Conclusions are given in Sec. VI.

\section{Model and nonlinear quantum regime}
We consider an OMS with modulated radiation-pressure coupling and mechanical spring constant, depicted in Fig.\,\ref{fig1}(a).
The modulation of the spring constant $k(t)$ induces a mechanical parametric amplification with frequency $2 \omega_{d}$, amplitude $\lambda$, and phase $\Phi_{d}$~\cite{Rugar1991,Szorkovszky2011Doherty,Szorkovszky2013Brawley}.
Then in a frame rotating with frequency $\omega_{d},$ the system Hamiltonian reads ($\hbar=1$)
\begin{eqnarray}
H_{\rm tot} &=& \omega_{c} a^\dag a+\Delta_m b^\dag b+\frac 12 \lambda(b^{\dag2} e^{-i \Phi_d} + b^2 e^{i \Phi_d})
\nonumber \\
 &&-\frac 12 g_{0} a^\dag a (b+b^\dag + b e^{-2i \omega_d t}+b^\dag e^{2i \omega_d t}),
 \label{e1}
\end{eqnarray}
where $a\;(a^\dag)$ and $b\;(b^\dag)$ are the annihilation (creation) operators of the cavity mode and mechanical mode, respectively.
The frequency detuning is $\Delta_m=\omega_m-\omega_d$ with the optical and mechanical resonant frequencies $\omega_c$ and $\omega_m$.

Here we assume that the mechanical oscillator is coupled to a squeezed vacuum reservoir with the center squeezing parameter $r_e$ and reference phase $\Phi_e$.
Including both optical and mechanical dissipations, the Lindblad superoperators read $\kappa D[a] \rho$ for the cavity damping and $\gamma(N +1) \mathcal{D}[b] \rho + \gamma N \mathcal{D}[b^\dag] \rho - \gamma M \mathcal{G} [b] \rho- \gamma {M}^{\ast} \mathcal{G} [b^\dag] \rho$ for the mechanical damping, where $D[o] \rho = o \rho o^\dag-\frac{1}{2}
(o^\dag o \rho + \rho o^\dag o),\mathcal{G} [o] \rho = o \rho o-\frac{1}{2}(o o \rho + \rho o o).$ Here $\kappa$ and $\gamma$ denote the photon and phonon decay rates, respectively. In addition, the expressions $N=\sinh ^2 (r_e),$ $M=\sinh(r_e) \cosh(r_e) e^{i \Phi_e}$ correspond to the effective thermal phonon number and the two-phonon correlation strength for the original phonon mode $b$~\cite{Breuer2002}.

\begin{figure}[htb]
\centering
\includegraphics[width=0.46\textwidth]{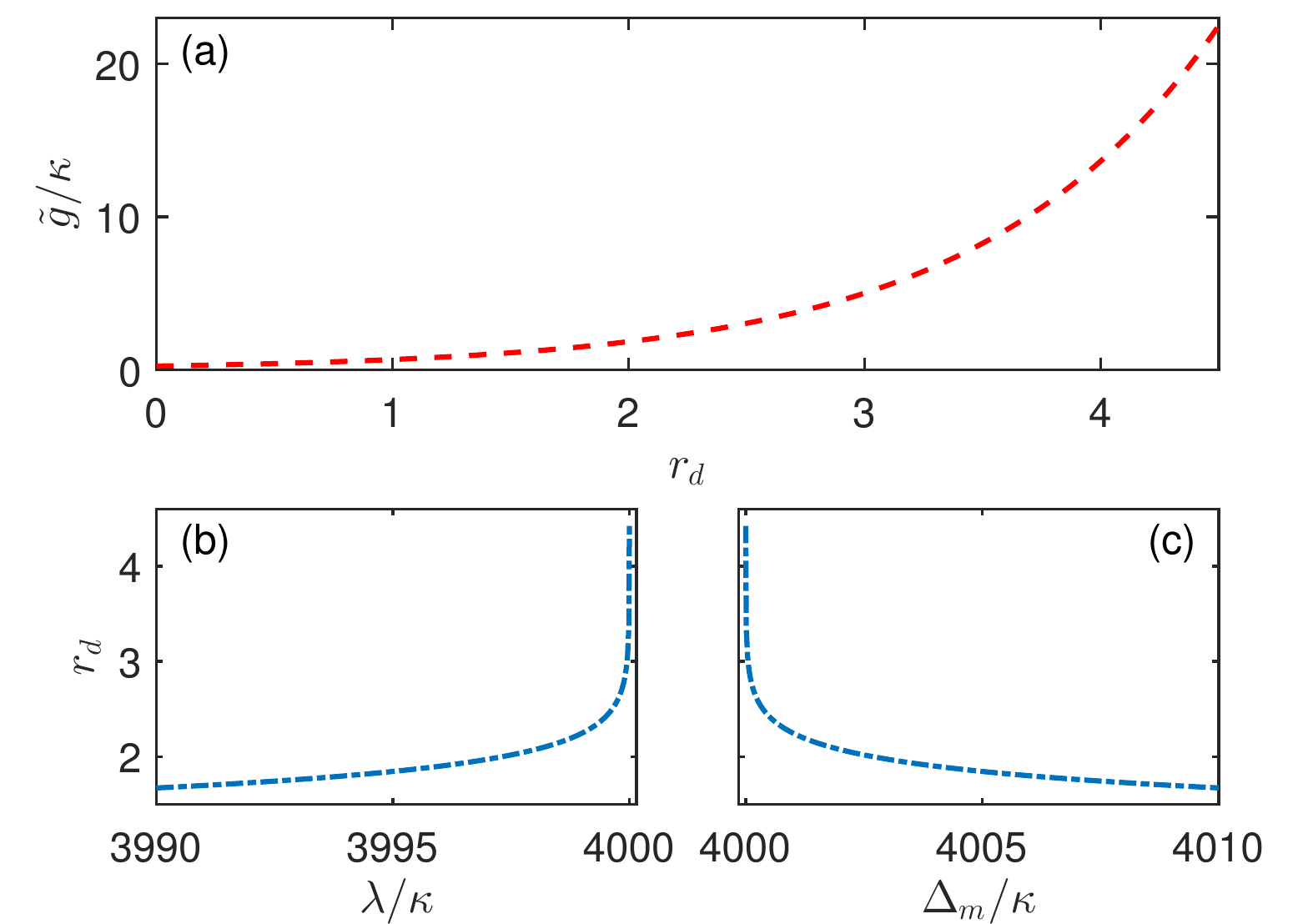}
\caption{(Color online) (a) The effective single-photon optomechanical coupling strength $\tilde{g}/ \kappa$ versus the squeezing parameter $r_d$.
The squeezing parameter $r_d$ versus (b) driving amplitude $\lambda/\kappa$ and (c) frequency detuning $\Delta_m/\kappa$.
The system parameters are scaled by the decay rate $\kappa$, i.e., (a) $g_0=0.5\kappa$, (b) $\Delta_m=4000\kappa$, and (c) $\lambda=4000\kappa$.}
\label{fig2}
\end{figure}

The quadratic part of $H_{\rm tot}$ can be diagonalized by introducing a squeezing transformation
$b=\tilde{b}\cosh (r_d) - \tilde{b}^\dag e^{-i \Phi_d} \sinh (r_d)$, with $r_d=\frac 14 \ln \frac {\Delta_m+\lambda}{\Delta_m-\lambda}$. Then, in terms of $\tilde{b}$, Hamiltonian (\ref{e1}) is expressed as
\begin{subequations}
\label{e2}
\begin{eqnarray}
H_{\rm tot} &=&\omega_c a^\dag a+\tilde{\omega}_{m} \tilde{b}^\dag \tilde{b}-\tilde{g} a^\dag a (\tilde{b}+\tilde{b}^\dag)+H_{\rm nr},
\\
\!\!\!\!\!\!H_{\rm nr}\!\!&=&\!\!-\frac 12 g_{0} a^\dag a\left\{\left[\tilde{b}\cosh ({r_d})\!\!-\!\!\tilde{b} ^\dag \sinh ({r_d})\right]\!\!e^{-2i \omega_d t}\!\!+\!\!{\rm H.c.}\right\},\nonumber
\\
\end{eqnarray}
\end{subequations}
where $\tilde{\omega}_{m}=\Delta_m /\cosh{(2 r_d)}$ is the transformed mechanical frequency, $\tilde{g}=\frac 12 g_{0} e^{r_d}$ is the enhanced optomechanical coupling, and the value of phase $\Phi_d$ is set to $\pi$.
Under the conditions $\omega_d \gg \tilde{\omega}_{m}$, $g_0\cosh(r_d)$, $g_0\sinh(r_d)$, the Hamiltonian $H_{\rm nr}$ becomes the term that oscillates with high frequencies, $2 \omega_d \pm \tilde{\omega}_{m}$, and it can be safely ignored under the RWA. Also, the validity of this approximation is manifested in the following section. Then a standard optomechanical Hamiltonian is obtained and given by
\begin{equation}
H_{\rm OMS}=\omega_c a^\dag a+\tilde{\omega}_{m} \tilde{b}^\dag \tilde{b}-\tilde{g} a^\dag a (\tilde{b}+\tilde{b}^\dag),
\label{e3}
\end{equation}
with a significantly enhanced single-photon optomechanical coupling strength $\tilde{g}$.
As shown in Fig.\,\ref{fig2}, a large squeezing parameter $r_d$ could be clearly obtained by adjusting the system parameters $\Delta_m$ or $\lambda$, which leads the realization of the effective single-photon strong-coupling regime, i.e., $\tilde{g}>\kappa$. In other words, our system could enter into the nonlinear quantum regime even when it is originally in the weak-coupling regime $g_0<\kappa$.

Including the dissipation caused by the system-bath coupling, the dissipative dynamics of the optomechanical system in terms of the squeezed mode is described by the master equation
\begin{eqnarray}
\label{e4}
\dot{\rho}&=&-i[H_{\rm tot},\rho]+\kappa \mathcal{D}[a] \rho + \gamma(\tilde{N}+1) \mathcal{D}[\tilde{b}] \rho
\nonumber \\
&&+ \gamma \tilde{N}\mathcal{D}[\tilde{b}^\dag] \rho - \gamma \tilde{M} \mathcal{G} [\tilde{b}] \rho - \gamma \tilde{M}^{*} \mathcal{G} [\tilde{b}^\dag] \rho,
\end{eqnarray}
where the Hamiltonian $H_{\rm tot}$ is given by Eq.\,(\ref{e2}) and $\tilde{N}$, $\tilde{M}$ denote the effective thermal noise and two-phonon correlation interaction, respectively, with the expressions of
\begin{subequations}
\label{e5}
\begin{align}
\tilde{N} = &\sinh ^2 (r_e) \cosh ^2 (r_d)+ \sinh ^2 (r_d) \cosh ^2 (r_e)
 \nonumber\\
&+\frac 12 \cos(\Phi) \sinh (2 r_e) \sinh (2 r_d)\label{e5a},
\\
\tilde{M} = &e^{i \Phi_d}[\cosh(r_e) \cosh(r_d)+e^{-i \Phi} \sinh(r_e) \sinh(r_d)]
\nonumber \\
&\times[\cosh(r_e) \sinh(r_d)+e^{i \Phi} \sinh(r_e) \cosh(r_d)].
\end{align}
\end{subequations}

Here the relative phase is $\Phi=\Phi_e-\Phi_d$. Note that, under the ideal parameter conditions $r_e=r_d$ and $\Phi=\pm n\pi$ $( n=1,3,5,... )$, the effective thermal occupancy of the bath for the squeezed mechanical mode $\tilde{b}$ and the two-phonon correlation strength can be suppressed completely, i.e., $\tilde{N}$, $\tilde{M}=0$. Contrarily, the effective thermal noise of mode $\tilde{b}$ will be amplified significantly when no squeezing is applied into the mechanical bath (i.e., $r_e=0$). For example, one can obtain that $\tilde{N}=\sinh^2(r_d)$, $\tilde{M}=\sinh(2 r_d)/2$, when the mechanical oscillator is initially in a vacuum bath, i.e., $r_e=0$.

It is necessary to suppress the effective thermal noise $\tilde{N}$ and two-phonon correlation interaction $\tilde{M}$ to observe the single-photon (or few-photon) quantum effects based on our proposal, as shown in the next section.
Specifically, in our proposal, we can obtain the single-photon strong-coupling regime by the simultaneous modulation of the mechanical spring constant and the optomechanical coupling. However, meanwhile, an unavoidable consequence is the amplified thermal noise induced by the mechanical parametric amplification, i.e., $\tilde{N}=\sinh^2(r_d)$, $\tilde{M}=\sinh(2 r_d)/2$, when the mechanical oscillator is initially in a vacuum bath.
Therefore, the quantum effects will be covered by the amplified mechanical thermal noise, even when the optomechanical interaction enters into the single-photon strong-coupling regime.

Contrarily, in our proposal, the quantum property of the system can be increased substantially with a bath, which is vanishingly squeezed under the ideal parameter condition $r_e=r_d$ and $\Phi=\pm n\pi$ $( n=1,3,5,... )$. Qualitatively, this result can be understood from the phase matching in Fig.\,\ref{fig1}(b)~\cite{Lv2015Wu}. The reservoir of the original mechanical mode $b$ is squeezed along the two quadrature axis with an angle $\Phi_e/2$, with a squeezing parameter $r_e$. In the basis of the squeezed mechanical mode $\tilde{b}$, this effect is cancelled by the squeezing (along axis $\Phi_d/2$) induced by the parametric amplification of $b$, when $\Phi_e-\Phi_d=\pm n\pi$ and $r_e=r_d$. That is, the squeezed vacuum reservoir (ellipse) of $b$ corresponds to an effective vacuum reservoir (circle) of $\tilde{b}$.

\begin{figure}[htb]
\centering\includegraphics[width=0.47\textwidth]{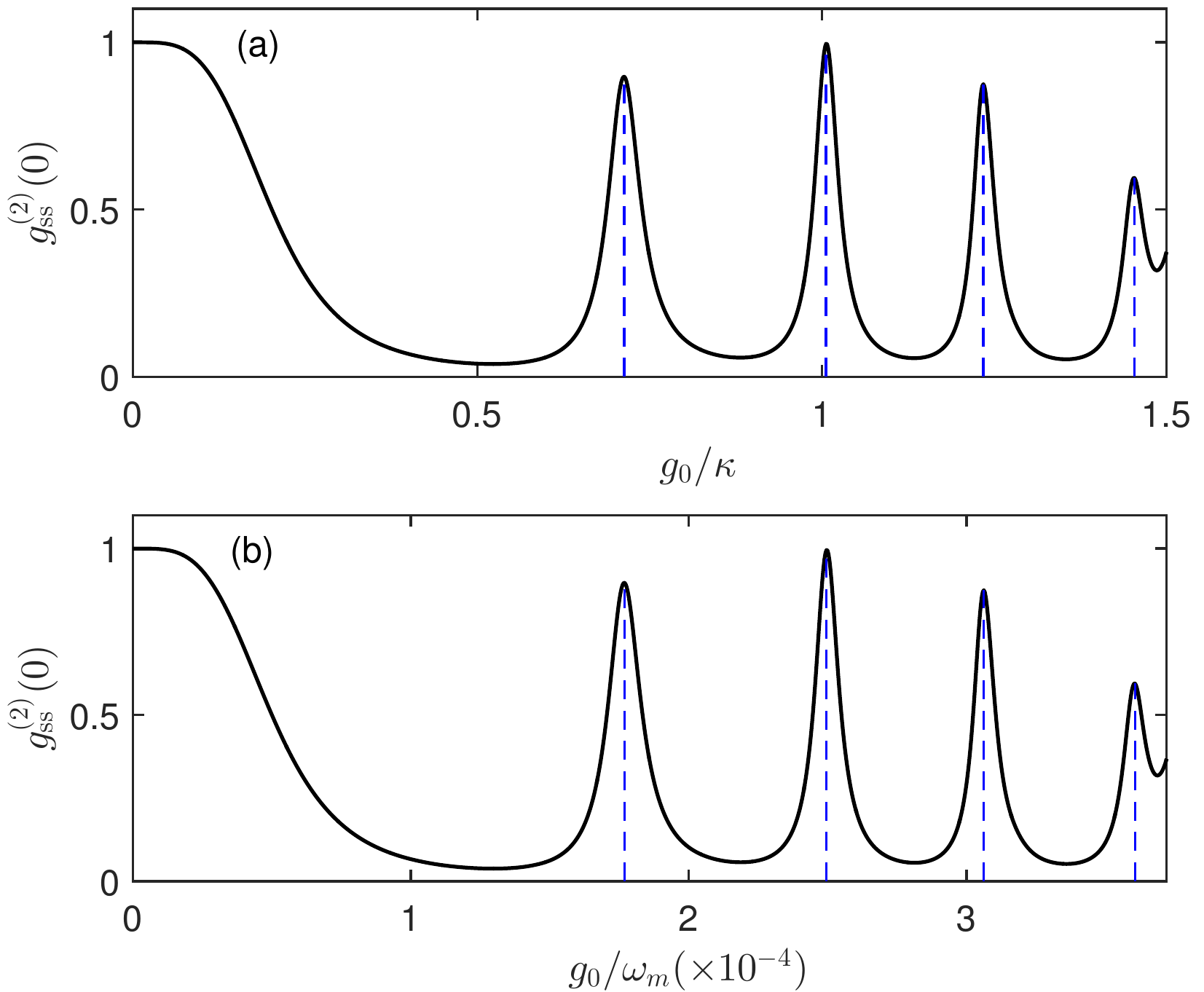}
\caption{(Color online) The steady-state equal-time second-order correlation function $g^{2}_{\rm{ss}}(0)$ versus the scaled coupling (a) $g_0/\kappa$ and (b) $g_0/\omega_{m}$ in the single-photon-resonance case $\Delta_c={\tilde{g}}^2/\tilde{\omega}_{m}$ with $H'_{\rm OMS}$ for the ideal parameter matching. The parameters we take are $\Delta_m=4000\kappa, \omega_d=30\kappa, \delta=0.02\kappa, \gamma=0.01\kappa, \varepsilon_{p}=0.1\kappa$, and $\Phi=\pi$. Here the symbol $\delta$ denotes the difference between the parameters $\Delta_m$ and $\lambda$ as $\delta=\Delta_m-\lambda$.
The relevant squeezing parameters are $r_e=r_d \approx 3.22$.
}
\label{fig3}
\end{figure}

\section{Photon blockade}
To exhibit the nonlinear quantum regime, we investigate the statistical properties of the cavity field, which are characterized by the equal-time second-order correlation function in the steady state
\begin{equation}
g^{2}_{\rm{ss}}(0) ={\rm{Lim}}_{t \rightarrow \infty }\frac{\langle a^\dag a^\dag a a \rangle(t)}{\langle a^\dag a \rangle ^2(t)} \nonumber
\end{equation}
and in the transient state
\begin{equation}
g^{2}(0)=\frac{\langle a^\dag a^\dag a a \rangle(t)}{\langle a^\dag a \rangle ^2(t)}. \nonumber
\end{equation}
In other words, the photon anti-bunching effect [$g^{2}_{\rm{ss}}(0)<1$] and even photon blockade [$g^{2}_{\rm{ss}}(0)\rightarrow0$] can be realized in the nonlinear quantum regime.

Then we drive the optical cavity with a weak probe field with frequency $\omega_p$ and amplitude $\varepsilon_p$ ($\varepsilon_p\ll\kappa$).
The Hamiltonian is $H_p=\varepsilon_p(a +a^\dag)$ in the frame rotating with the probe frequency $\omega_p$.
We calculate the steady-state equal-time second-order correlation function $g^{2}_{\rm{ss}}(0)$ by numerically solving the master equation in Eq.\,(\ref{e4}),
\begin{figure}[htb]
\centering
\includegraphics[width=0.47\textwidth]{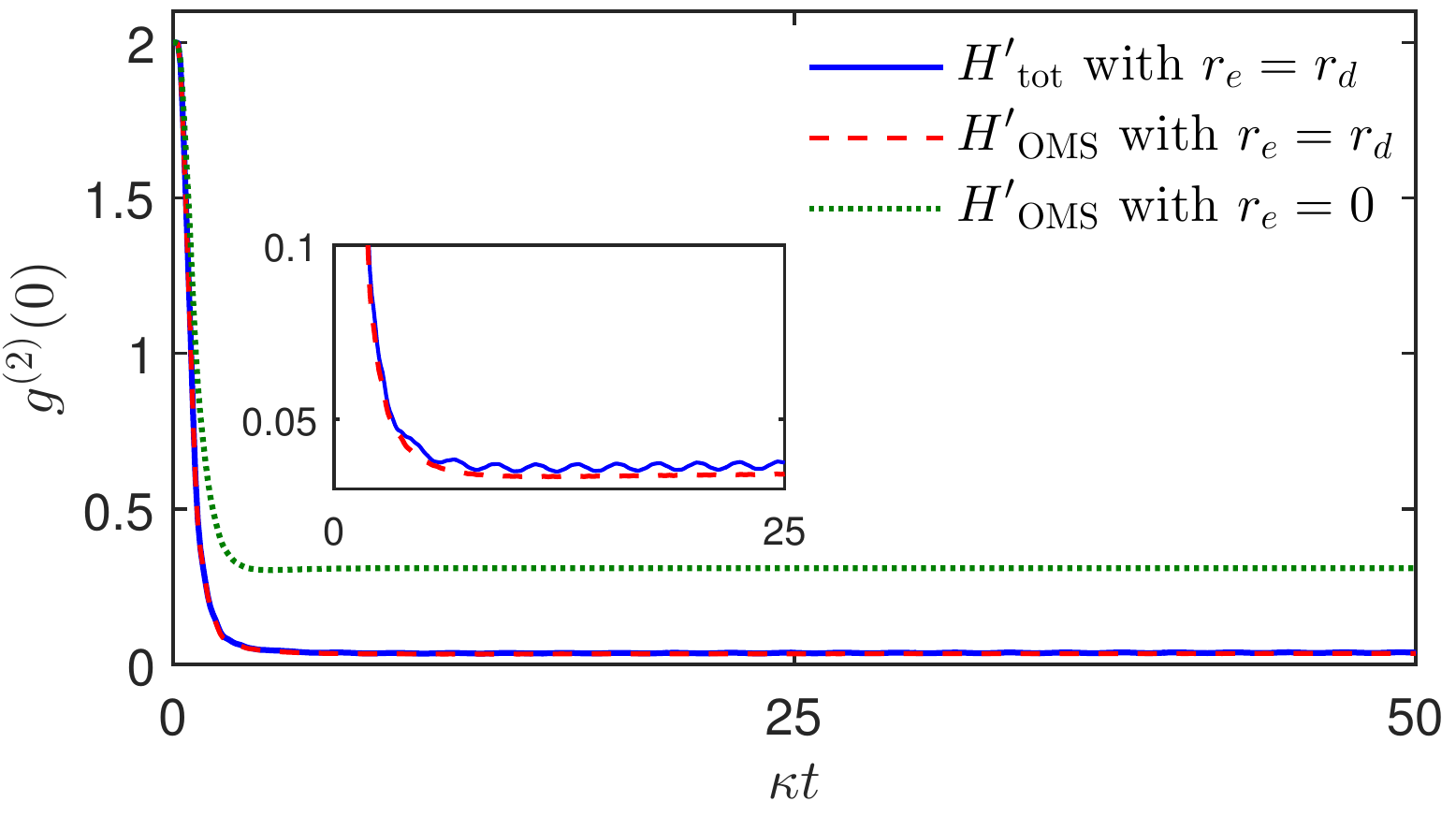}
\caption{(Color online) The equal-time second-order correlation function $g^{2}(0)$ versus the scaled time $\kappa t$ in the single-photon-resonance case $\Delta_c={\tilde{g}}^2/\tilde{\omega}_{m}$.
The cavity and mechanical modes are initially in a thermal state and vacuum state, respectively. All the curves are obtained by numerically calculating Eq.\,(\ref{e4}) with $H'_{\rm tot}$ or $H'_{\rm OMS}$.
 The parameters are the same as that in Fig.\,\ref{fig3} except for $g_0=0.5\kappa$. Here the green dotted line (i.e., $r_e=0$) actually corresponds to the case in which the mechanical oscillator is initially in a vacuum bath and an amplified thermal noise $\tilde{N}$=$\sinh^2(r_d)$ is induced for the considered squeezed mode $\tilde{b}$.}
\label{fig4}
\end{figure}
but with the effective system Hamiltonian
\begin{figure}[htb]
\centering
\includegraphics[width=0.48\textwidth]{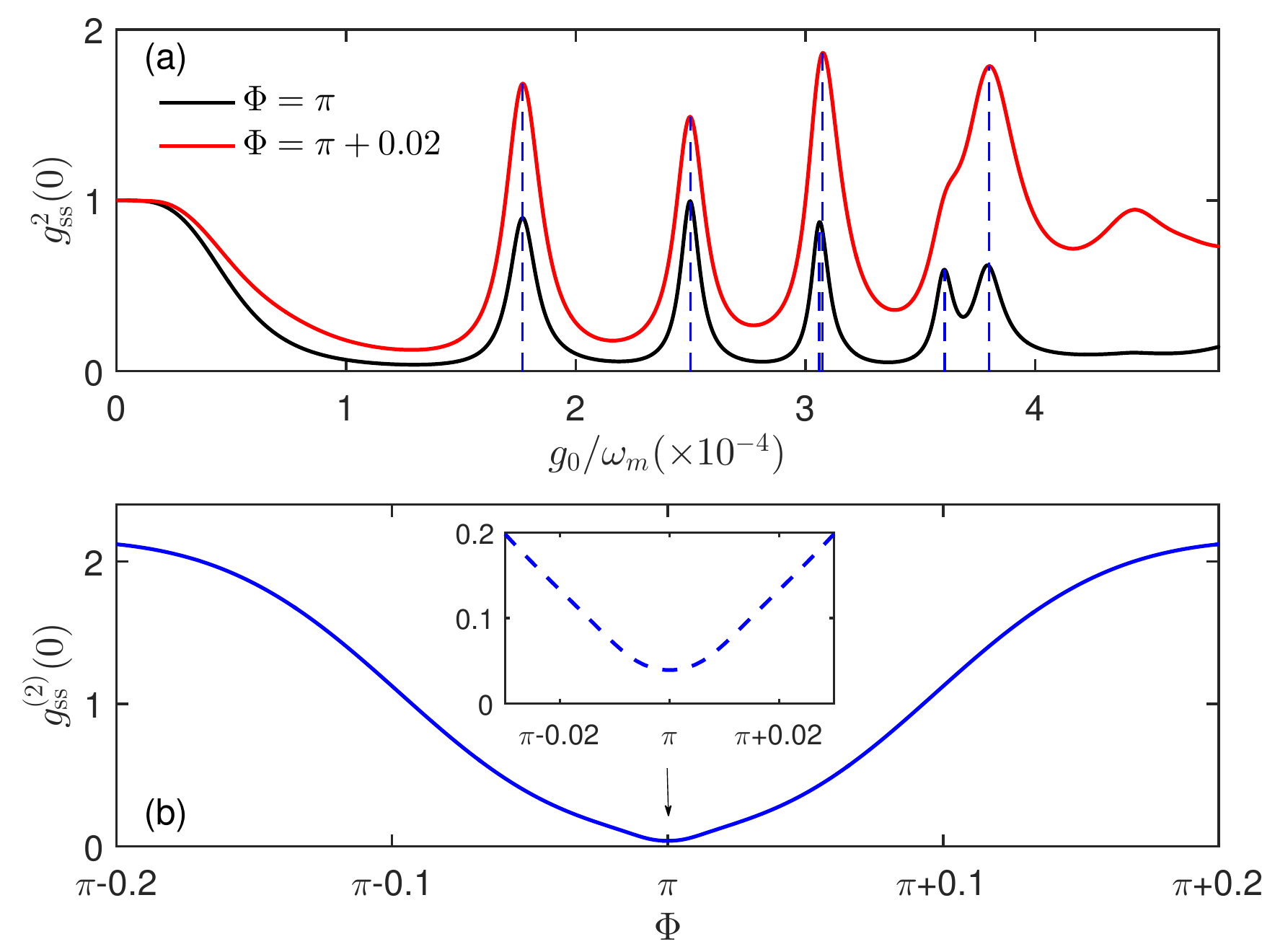}
\caption{(Color online) The steady-state equal-time second-order correlation function $g^{2}_{\rm{ss}}(0)$ versus (a) $ g_0/\omega_{m}$ with different $\Phi$,
(b) $\Phi$ for the case of single-photon resonance $\Delta_c={\tilde{g}}^2/\tilde{\omega}_{m}$.
The system parameters are the same as that in Fig.\,\ref{fig3} except for (b) $g_0=0.5\kappa$.}
\label{fig5}
\end{figure}
$H'_{\rm OMS}=\Delta_c a^\dag a+\tilde{\omega}_{m} \tilde{b}^\dag \tilde{b}-\tilde{g} a^\dag a (\tilde{b}+\tilde{b}^\dag)+H_p$ and $\Delta_c =\omega_c-\omega_p$.
In Fig.\,\ref{fig3}, we show $g^{2}_{\rm{ss}}(0)$ as a function of the scaled coupling $g_0/\kappa$ and $g_0/\omega_{m}$, respectively, in the case of single-photon resonance (i.e., $\Delta_c={\tilde{g}}^2/\tilde{\omega}_{m}$) and the ideal parameter matching conditions [i.e., $r_e=r_d$ and $\Phi=\pm n\pi$ $(n=1,3,5,...)$]. It is shown that the strong photon anti-bunching effect $g^{2}_{\rm{ss}}(0)<1$ and even the photon blockade $g^{2}_{\rm{ss}}(0)\rightarrow0$ can be obtained in the weak coupling regime $g_0<\kappa$ and $g_0\ll\omega_m$. This clearly demonstrates the achievement of nonlinear quantum regime in a weakly coupled OMS.
Moreover, the photon tunneling effect is also exhibited at $\tilde{g}/\tilde{\omega}_{m}=\sqrt{m/2}$ $(m=1,2...)$, corresponding to the occurrence of two-photon resonant effect (see the dashed blue lines in Fig.\,\ref{fig3}).

To check the validity of the RWA applied in the above calculations, in Fig.\,\ref{fig4} we numerically present the evolution of $g^{2}(0)$ with the total Hamiltonian of the system including the probe field, i.e., $H'_{\rm tot}=H'_{\rm OMS}+H_{\rm nr}$. It clearly shows that the approximate result corresponding to $H'_{\rm OMS}$ agrees well with the exact numerical calculations using $H'_{\rm tot}$. The slightly oscillating feature in the case of $H'_{\rm tot}$ comes from the high-frequency-oscillation term $H_{\rm nr}$ (see the insert of Fig.\,\ref{fig4}). Moreover, in the case of $r_e=0$, while the photon anti-bunching effect is maintained, the photon blockade is destroyed by the amplified phonon noise (see the green dotted line of Fig.\,\ref{fig4}). This corresponds to the case of the mechanical oscillator being initially in a vacuum bath.
\begin{figure}[htb]
\centering
\includegraphics[width=0.47\textwidth]{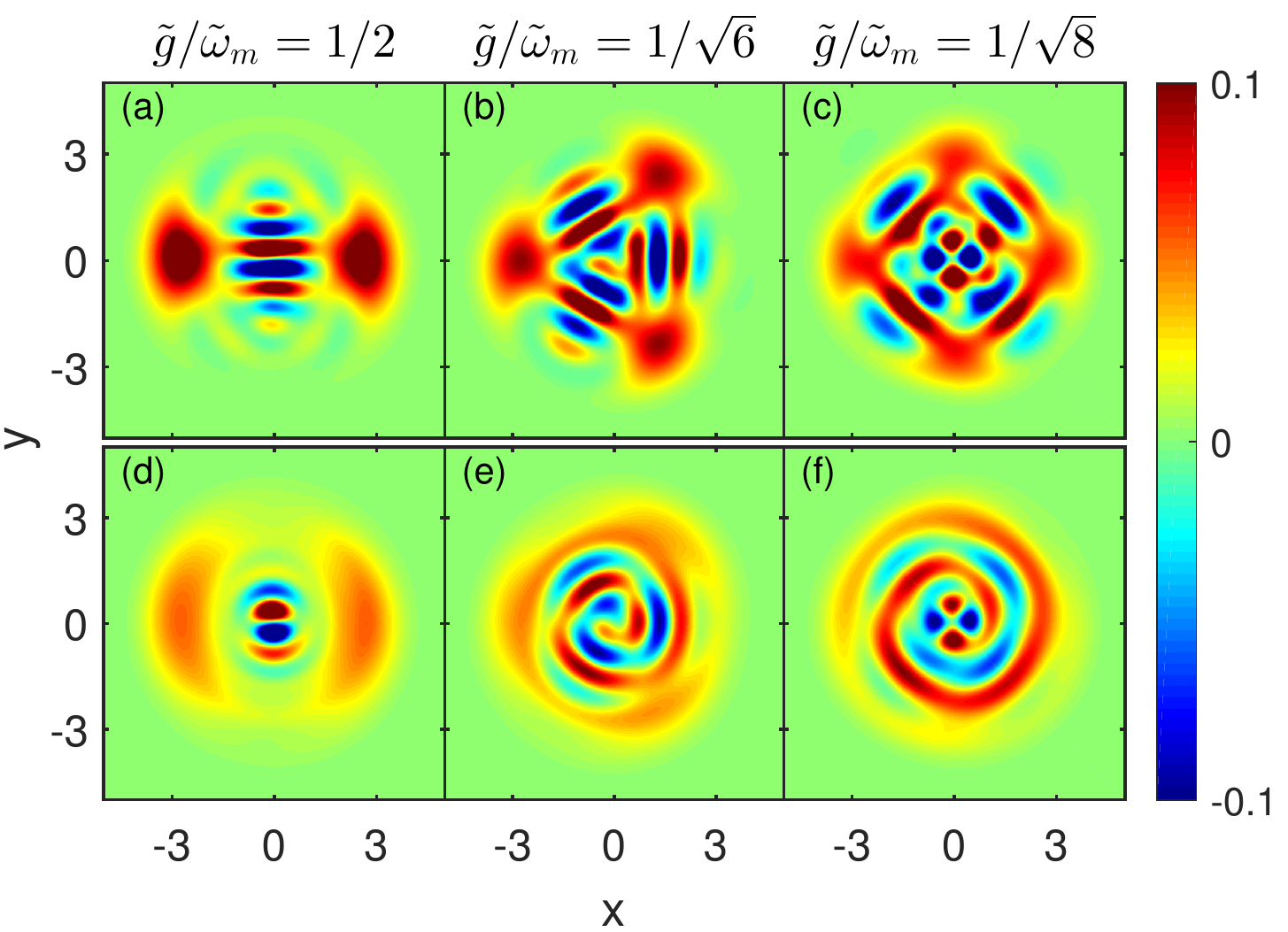}
\caption{(Color online) Wigner function of the cavity field at time $t=2 \pi/\tilde{\omega}_{m}$ for different $\tilde{g}/\tilde{\omega}_{m}$ when (a)-(c) $r_e=r_d$ and (d)-(f) $r_e=0$.
The case of $r_e=0$ is the same as for the green dotted line of Fig.\,\ref{fig4}.
The quadrature variables are $ x=(a +a^\dag)/2,y=-i(a-a^\dag)/2$. The cavity field and the mechanical mirror are initially in coherent states $|\alpha\rangle$, $|\beta\rangle$ with the amplitudes $\alpha=\beta=2$.
Other parameters are the same as that in Fig.\,\ref{fig3} except for (a), (d) $\kappa /\omega_{m}=3.16\times10^{-5}$, $\gamma /\kappa=10^{-2}$, $g_0 /\omega_{m}=1.26\times10^{-4}$, (b) (e) $g_0 /\omega_{m}=1.03\times10^{-4}$, and (c), (f) $g_0 /\omega_{m}=0.89\times10^{-4}$.}
\label{fig6}
\end{figure}

\begin{figure}[b]
\centering\includegraphics[width=0.48\textwidth]{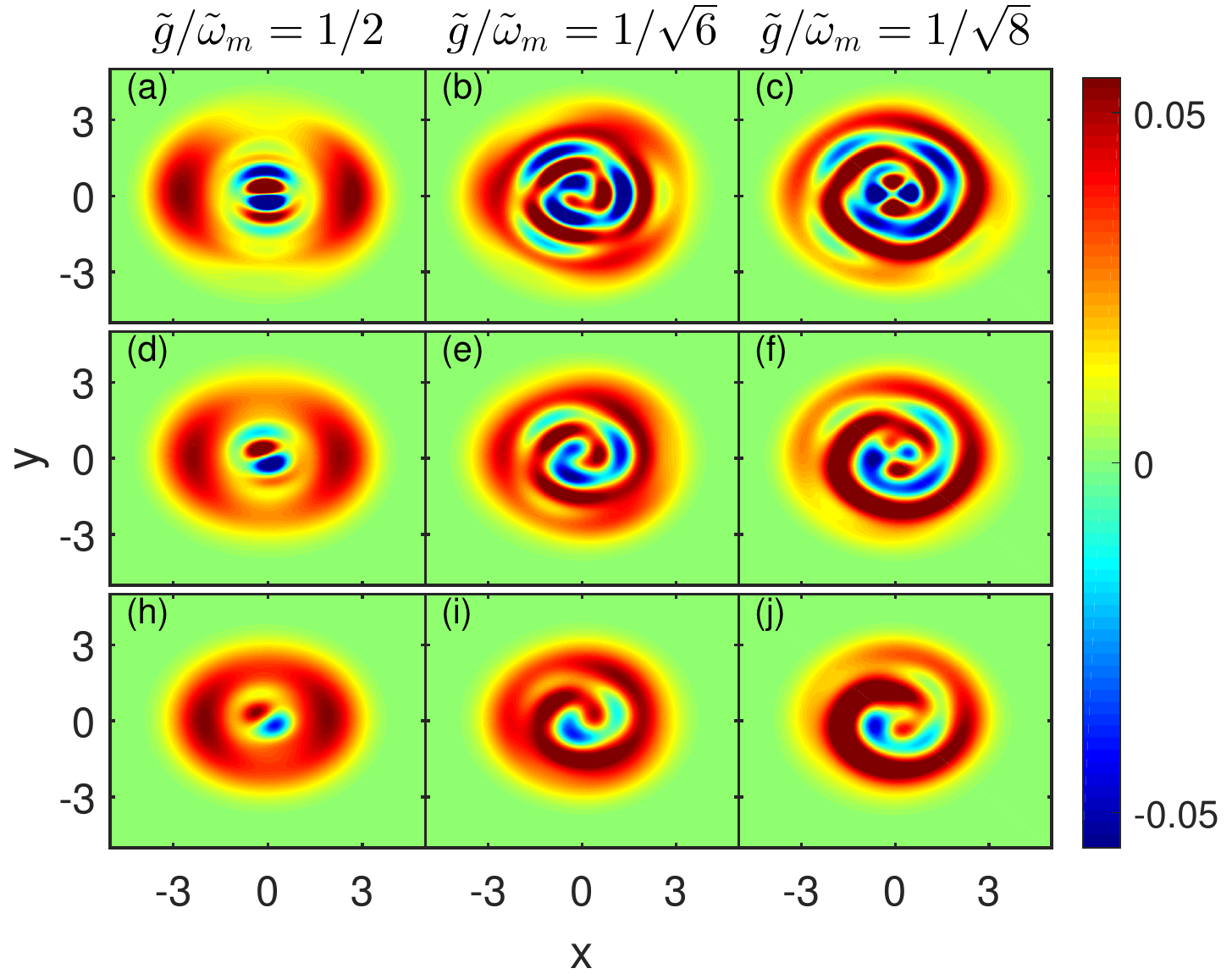}
\caption{(Color online) Wigner function of the cavity field at time $t=2\pi/\tilde{\omega}_{m}$ for different $\tilde{g}/\tilde{\omega}_{m}$.
The system parameters are the same as in Fig.\,\ref{fig6} except for $r_e=0$ and
(a)-(c) $\kappa /\omega_{m}=3.16\times10^{-5}$, (d)-(f) $\kappa/\omega_{m}=1.58\times10^{-4}$, (h)-(j) $\kappa/\omega_{m}=3.16\times10^{-4}$, (a), (d), (h) $g_0/\omega_{m}=1.26\times10^{-4}$,
(b), (e), (i) $g_0/\omega_{m}=1.03\times10^{-4}$, and (c), (f), (j) $g_0/\omega_{m}=0.89\times10^{-4}$.}
\label{fig7}
\end{figure}

The above results show that a squeezed vacuum environment with optimal squeezing strength $r_e$ is required to observe photon blockade.
The influence of the squeezing phase $\Phi$ on the photon statistical properties is presented in Fig.\,\ref{fig5}. It shows that the photon blockade occurs in the vicinity of the phase matching $\Phi=\pi$, at which the thermal noise for the mechanical mode is completely suppressed. The system is thermalized by the amplified phonon noise $\tilde{N}$ when $\Phi$ deviates too much from $\pi$. In this case, the two-photon tunneling is enhanced by the parametric-amplification-induced two-phonon-correlation effects (i.e., $\tilde{M}\neq0$), while the tunneling peaks have a very slight shift.
In a short summary, although the nonlinear quantum regime characterized by $g^{2}_{\rm{ss}}(0)<1$ is robust to the amplified phonon noise $\tilde{N}$, photon blockade is highly sensitive to $\tilde{N}$.
This requires that the stable squeezing effects should be applied into the mechanical bath during the interaction time $\kappa t \approx 10$, which is experimentally feasible by employing a driving laser with stable frequency and phase to induce this squeezing effects~\cite{Jiang2011,Fortier2011}. Specifically, as shown in Fig.\,\ref{fig4}, $g^{2}(0)$ approaches a steady value when $\kappa t \approx 10$, i.e., for $\kappa $=0.1 MHz, $t\approx 100 \,\mu s$.

\section{Nonclassical states of cavity field}
Normally, strong Kerr nonlinearities could be induced when the OMS enters into the nonlinear quantum regime, which ultimately leads to the generation of nonclassical states (i.e., Schr\"{o}dinger cat states) of the cavity field ~\cite{Mancini1997,Bose1997,Marshall2003,Lv2013Zhang}. This has important applications in quantum information science. Now we show the generation of Schr\"{o}dinger cat states with a relatively weak optomechanical coupling (i.e., $g_0 \ll \omega_m$) based on our proposal.

In Figs.\,\ref{fig6} and \ref{fig7}, we plot the Wigner function of the cavity field at time $t=2 \pi/\tilde{\omega}_{m}$ for various values of the scaled coupling $\tilde{g}/\tilde{\omega}_{m}$ based on Eq.\,(\ref{e4}) with the Hamiltonian $H_{\rm OMS}$ (omitting the free evolution of the cavity mode $a$ in the interaction picture). First, it shows that the two-, three- and four-component Schr\"{o}dinger cat states could be obtained based on our proposal. This is because our system could enter into an effective quantum nonlinear regime (i.e., $\tilde{g}>\kappa$ and $\tilde{g}\sim\tilde{\omega}_m$) even when it is in the relatively weak optomechanical coupling
case (i.e., $g_0\ll\omega_m$). Second, the generated cat states are robust against the amplification-induced phonon noise $\tilde{N}$ of mode $\tilde{b}$, comparing the case of $\tilde{N}=0$ [Figs.\,\ref{fig6}(a)-\ref{fig6}(c)] with $\tilde{N}=\sinh^2(r_d)$ [Figs.\,\ref{fig6}(d)-\ref{fig6}(f)]. In other words, to prepare a nonclassical state of the cavity field, our proposal does not require a mechanical bath with elaborate designing. This is essentially due to the small mechanical damping rate $\gamma$ in the typical OMS. Finally, the generation of cat states is sensitive to the optical decay, as shown in Fig.\,\ref{fig7}. Specifically, the negativity of Wigner function only shows a very narrow range and the shape of multi-component Shr\"{o}dinger cat states almost disappears when $\kappa/\omega_{m}\geq3.16\times10^{-4}$. It shows that the quantum property of the cavity field is destroyed by the strong decoherence effects when the optical decay rate is too large.

\section{Discussions}
In this section, let us discuss the experimental prospect of our proposal. First, it should be clearly mentioned that our proposal is not doable with current technology since a relatively large optomechanical coupling strength (i.e., $g_0=0.5\kappa$) has been used in our calculations; e.g., Fig.\,\ref{fig2}(a). However, in principle, our proposal is also applicable to the case of a smaller $g_0$ by employing a larger squeezing parameter $r_d$ in the critical parameter regime $\Delta_m\rightarrow\lambda$, as shown in Figs.\,\ref{fig2}(b) and \ref{fig2}(c). Note that the critical parameter regime, where $\Delta_m$ approaches $\lambda$, is feasible with current laser technologies~\cite{Jiang2011,Fortier2011}, although there also exists an experimental challenge. Moreover, our proposal is also useful for relaxing the parameter condition of implementing single-photon quantum-processing with an OMS, because the implementation of single-photon quantum-processing based on the previous proposal~\cite{Rabl2011,Kronwald2013Ludwig,Liao2013Law} usually requires that $g_0$ is much larger than $\kappa$ and approaches $\omega_m$.

Second, our proposal is general and applicable to the OMS system in the optical wave range or electromechanical system in the microwave range. Choosing the ultrahigh Q toroid microcavity as an example~\cite{Armani2003}, to implement our proposal in the future, the system parameters could be taken as $\kappa$=0.1 MHz, $\gamma=0.01\kappa=10^{-3}$ MHz, $g_0=0.5\kappa=0.05$ MHz, $\Delta_m=4000\kappa$=400 MHz, $\delta=\Delta_m-\lambda=0.02\kappa=2\times10^{-3}$ MHz, and $\omega_d=30\kappa$=3 MHz.

Finally, in our proposal, a squeezed mechanical bath is needed to reach the enhanced quantum effects. In principle, it could be realized by introducing an ancillary cavity mode, which is initially in a squeezed optical environment~\cite{Gu2013Li,Wang2013Clerk,Nunnenkamp2014Sudhir} and adiabatically follows the mechanical dynamics. Specifically, we consider an ancillary cavity mode $f$ (with resonant frequency $\omega_{f}$ and $\omega_f-\omega_c\gg\omega_m$) coupled to the mechanical oscillator $b$. Similar to the case of optical mode $a$, a modulated radiation-pressure coupling between $f$ and $b$ is obtained, i.e., $g_{f}\cos(\omega_{d} t)$, when the period modulation is applied on the mechanical spring constant. Under the condition of strong driving for the ancillary cavity mode $f$, the resonant optomechanical interaction between modes $f$ and $b$ will be linearized and becomes $G (f+f^\dag) (b+b^\dag)$ with $G=\frac{1}{2}g_f \sqrt{n_f}$ ($n_f$ is the mean photon number in the cavity of the ancillary mode $f$).
Under the assumption that the decay rate of the ancillary mode $f$ is much larger than the linearized optomechanical interaction, i.e., $\kappa_f \gg G$, we can adiabatically eliminate $f$ in the red-detuned driving case.
In the resolved-sideband regime $\omega_{m} \gg \kappa_f$, the mechanical oscillator will adiabatically follows the dynamics of mode $f$~\cite{Vitali2007Gigan} and one has
$b \simeq  [\kappa_f/(2 i G)] f - [\sqrt{\kappa_f}/(i G)] f_{{\rm in}}$, where $f_{{\rm in}}$ is the input noise of the ancillary mode $f$. Moreover, the ancillary mode $f$ could be in a squeezed vacuum environment by interacting with a broadband-squeezed vacuum field generated by an optical parametric amplification~\cite{Lv2015Wu}. Then this squeezed vacuum bath could be transferred to the mechanical oscillator under the above adiabatic condition in principle~\cite{Gu2013Li}. Note that this approach by using the related ancillary cavity to effectively change the mechanical bath has already been applied in cavity optomechanics, such as achieving the enhanced mechanical damping~\cite{Wang2013Clerk,Nunnenkamp2014Sudhir}.

\section{Conclusion}
We have studied the quantum property of a weakly coupled OMS, where the optomechanical interaction as well as the mechanical spring constant are periodically modulated.
We have shown that our system could effectively enter into the nonlinear quantum regime by the spring-modulation-induced mechanical parametric amplification and the coupling-modulation-induced resonance interaction.
Specifically, the optomechanical interaction can be enhanced into the single-photon strong-coupling regime in an original weakly coupled OMS.
The generation of photon blockade and nonclassical states of the cavity field are demonstrated in the weakly coupled OMS.
This study provides a promising route to reach the nonlinear quantum regime of an OMS with currently available technology, and has potential applications in modern quantum science.

\begin{acknowledgements}
This work is supported by the National Key Research and Development Program of China grant No. 2016YFA0301203-02,
the National Science Foundation of China (Grant No. 11374116, No. 11574104 and No. 11375067).
\end{acknowledgements}

\end{document}